\documentstyle[12pt,fleqn]{article}
\begin{document}
\newcommand{\be}{\begin{equation}}
\newcommand{\ee}{\end{equation}}
\newcommand{\ds}{\displaystyle}
\renewcommand{\theequation}{\arabic{section}.\arabic{equation}}
\hfill{\normalsize ITP-95-9E}

\hfill{\normalsize March 1995}

\begin{center}
{\Large \bf Quark-exchange Effects in a Deuteron Breakup at
Intermediate Energy}\\
\vspace{1cm}
{\large A.~P.~Kobushkin and A.~I.~Syamtomov}\\[.3cm]
{\large \it N.N.Bogolyubov Institute for Theoretical Physics, \\
National Academy of Sciences of Ukraine, Kiev 143, Ukraine
}\\[0.5cm]
and\\[0.5cm]
{\large L.~Ya.~Glozman}\\[.3cm]
{\large\it Institute of Theoretical Physics, University of Graz,
A-8010 Graz, Austria\\
Alma-Ata Power Engineering Institute, 480013 Alma-Ata, Kazakhstan}
\end{center}
\date{}
\vspace{.5cm}
\begin{abstract}
Microscopical approach to a deuteron breakup at high and
intermediate energies is proposed. We show that the quark exchange effects,
resulting from the full asymmetry of the $6q$-deuteron wave function with
respect to the pair permutations of quark variables, strongly affect the
proton momentum distribution in the deuteron, as well as the
polarization observables of inclusive deuteron breakup, when the
``internal momentum'' in the deuteron is of order of a few hundreds MeV/c.
\end{abstract}
\newpage

\section{Introduction}

During the last decade the breakup of relativistic nonpolarized and polarized
deuterons on hydrogen/nuclear target have been a subject of experimental
and theoretical studies. They, as well as studies of $ed$-scattering at
high $Q^2$, bring light on many aspects of the nucleon-nucleon interaction
and the deuteron structure at short distances, including relativistic
effects, effects of non-nucleon degrees of freedom, {\em etc}. The
deuteron breakup experiments, we are discussing in this paper, include
measurements of differential cross-section,
$E_p \frac{\ds d^3\sigma}{\ds d\vec p}$, and polarization observables
(the tensor analyzing power, $T_{20}$, and the polarization transfer
coefficient, $\kappa_0$) of the $(d,p)$ reaction with the final proton
registered at zero angle and the momentum
$p \ \geq \ \frac{\ds 1}{\ds 2} \times  (\mbox{\rm deuteron lab. momentum})$
at deuteron beam energy of a few GeV.

We think that there are at least three kinematical regions
where different physical pictures determine the reaction.
At the region of the ``internal
momentum'' in the deuteron, $k\leq 0.1$ GeV/c, the relevant degrees of
freedom in the deuteron are nucleons and mesons. Here the
nonrelativistic impulse approximation (IA) improved by the rescattering
effects and the final state interactions, both calculated with
 the standard
two-nucleon deuteron wave functions, seems to be valid. At the region of
very high $k$ (according to estimations of \cite{Kob93} $k\geq 1$ GeV/c)
one has to provide a smooth transition to the perturbative QCD.

In the present paper we discuss the deuteron
breakup at the intermediate region ($k$ between 0.1 and 1 GeV/c) where the
nucleons in the deuteron are assumed to loose gradually their
individuality, but the pure perturbative QCD does not yet work. At this
region the deuteron structure can be described in terms of the constituent
quarks and the chiral mesons within the dynamical scheme similar to that
used in the baryon spectroscopy  \cite{GlozRis}.
Then one should note that the Pauli principle at the constituent
quark level gives rise to a number of the
short-range baryon-baryon components
($N\!N^{\star}$, $N^{\star}\!N$, $N^{\star}\!N^{\star}$) of the deuteron
wave function \cite{GlNeOb,Glozm}.

However, instead of the usual Fock probabilities for the baryon-baryon
components, which are applicable only for the structureless (or well
separated baryons), one should use a language of "effective numbers"
\cite{Glozm} developed in nuclear cluster physics \cite{Nemetz}.

In this work we shall restrict ourselves to a simplified model, in which
the quark antisymmetrization effects are taken into account for the
deu\-te\-ron $S$-wave only. Due to the centrifugal effect the
quark-exchange contribution for the $D$-wave is estimated to be three
orders of magnitude less than the one for the $S$-wave \cite{Glozm}.
Besides, we do not consider here
a delicate question of the relativistic internal momentum $k$ in the
deuteron: we use it in the same way as for the equal-mass $NN$ configuration
in the framework of the light-cone dynamics (``minimal relativistic
prescription''). The $3q\!\! \times \!\! 3q$ decomposition of the deuteron
six-quark wave function contains, apart from the equal-mass $NN$
configuration, a set of configurations, $NN^{\star}$, whose ``constituents''
($N$ and $N^{\star}$, respectively) have different masses. Meanwhile,
the main result of this work is that the quark exchanges between $3q$
clusters in the deuteron give a visible contribution in the differential
cross-section and polarization observables of the $(d,p)$ breakup in the
intermediate region, and have to be taken into consideration.

The outline of the paper is organized as follows. In Sec.~2 we consider
a decomposition of the deuteron six-quark
wave function into $3q\!\! \times \!\!3q$
clusters  and how it modifies a matrix element of the deuteron
breakup. In Sec.~3 expressions for the proton momentum distribution in
the deuteron and polarization observables are obtained. The results of
numerical calculations and conclusions are given in Sec.~4.

\begin{center}
\unitlength=0.80mm
\linethickness{0.4pt}
\begin{picture}(93.00,45.0)(20,50)
\put(35.00,58.92){\line(1,0){58}}
\put(35.00,59.78){\line(1,0){25}}
\put(60.00,59.78){\line(1,2){10}}
\put(72.00,80.86){\circle*{4.35}}
\put(42.00,89.03){\line(4,-1){31.09}}
\put(74.00,82.15){\line(5,3){12.97}}
\put(73.00,80.00){\line(2,-1){13.42}}
\put(74.00,80.86){\line(1,0){15}}
\put(87.00,92.90){\line(2,-5){5}}
\put(87.00,68.00){\line(2,5){5}}
\put(37,62){\makebox(0,0)[lb]{d}}
\put(90,65){\makebox(0,0)[lt]{$p'$}}
\put(45,90){\makebox(0,0)[lb]{$^1H$}}
\put(96,80){\makebox(0,0)[cc]{X}}
\put(60,70){\makebox(0,0)[cc]{B}}
\end{picture}
\end{center}
\begin{center}
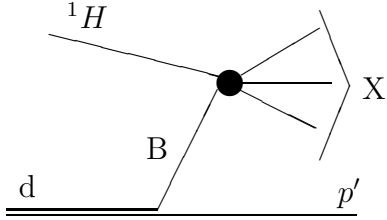
{Fig. 1. Deuteron inclusive breakup in spectator
approximation.}
\end{center}

\section{$NN$ and $NN^{\star}$ components of the deu\-te\-ron six-quark
wave function}

The deuteron wave function, considered at small internucleon distances
as a six-quark ($6q$) object, was shown
\cite{GlozKuch88,Obukh88,Obukh90} to be qualitatively
equivalent to the Resonating Group Method (RGM) wave
function
\begin{equation} \psi^{d}(1,2,\ldots,6)=
\hat{A}\left\{\varphi_{N}(1,2,3)
\varphi_{N}(4,5,6)\chi ({\bf r})\right\},\label{rgm}
\end{equation}
where $\hat{A}=\frac{1}{\sqrt{10}}(1-9\hat{P}_{36})$ is a quark
antisymmetrizer and $\varphi_{N}(1,2,3)$ and
$\varphi_{N}(4,5,6)$ are the wave functions of the  nucleon
three quark ($3q$) clusters; $\chi ({\bf r})$ is the RGM distribution
function.

The presence of the quark antisymmetrizer $\hat{A}$ in expression
(\ref{rgm}) is the main difference between the microscopical quark and
the meson-nucleon points of view on the deuteron structure at short
distances. In the absence of the quark antisymmetrizer the RGM
distribution function would coincide, up to the renormalization effects
at short range, with the
conventional deuteron wave function.

Due to the antisymmetrizer the deuteron wave function (\ref{rgm}), being
decomposed into $3q \times 3q$ clusters, includes, apart from the
standard $pn$ component, the nontrivial $NN^{\star}$, $N^{\star}N$
and $N^{\star}N^{\star}$ components which respond to all possible
nucleon resonance states, $N^{\star}$ (see, {\em e.g.},
\cite{Glozm}). So, in the spectator approximation (Fig.1)
the reaction matrix element is
\begin{equation}
M_{dp\rightarrow p^{\prime}X} =\sqrt{2} \sum_{B,M_{B}}
\psi^{d}_{Bp^{\prime}}({\bf k}_{p^{\prime}})
\Gamma_{Bp\rightarrow X},
\label{me}
\end{equation}
where the summation over $B$ includes $pn$ and all $pN^{\star}$
configurations of the deuteron; $M_B$ is  $z$-projection of $B$-baryon
spin. The other notations in (\ref{me}) are as follows:
$\Gamma_{Bp\rightarrow X}$ is the amplitude of the reaction
$B+p\rightarrow X$:
\begin{eqnarray}
\Gamma_{pB\rightarrow X} &\!\!\!=\!\!\!&\int \ \prod_{i=1}^{3}
\prod_{j=7}^{9}\ d^{3}r_{i}
\ d^{3}r_{j}\ e^{-i({\bf k}_{X}+{\bf k}_{p^{\prime}}-{\bf k}_{p})
{\bf r}_{X}} \nonumber \\
 & &\hskip -0.5in
\times\langle\psi_{X}(1,2,3,7,8,9)|\hat{\Gamma}(1,2,3,7,8,9)|
\varphi_{B}(1,2,3)\varphi_{p}(7,8,9)\rangle ,
\label{Gam}\\
{\bf r}_{X} &\!\!\!=\!\!\!& \frac{1}{6}\left(\sum_{i=1}^{3}  {\bf r}_{i} +
\sum_{i=7}^{9}  {\bf r}_{i}\right) ,
\label{rx}
\end{eqnarray}
and $\varphi_{B}(1,2,3)$ is a wave function of the $3q$ cluster
of the baryon $B$; $\varphi_{N}(7,8,9)$ is the same for the target proton;
${\bf r}_{i}$ stands for the $i$-th quark coordinate and
\be
\psi_{Bp^{\prime}}^{d}({\bf k}_{p^{\prime}}) =
\frac{1}{(2\pi)^{3/2}} \int \ d^{3}r\ e^{i{\bf k}_{p^{\prime}}{\bf
r}} \psi_{Bp^{\prime}}^{d}({\bf r}), \\
\label{Fourier}
\ee
is a Fourier transformation of the overlap,
$\psi_{Bp^{\prime}}^{d}({\bf r})$, between the $6q$ wave function (\ref{rgm})
of the deuteron and the $pB$ wave function; $\bf r$ stands
for the relative coordinate between the target and the baryon $B$:
\be
\psi_{Bp^{\prime}}^{d}({\bf r}) = \left(\frac{6!}{3!3!2}
\right)^{1/2}
\langle\varphi_{B}(1,2,3)\varphi_{p^{\prime}}(4,5,6)
|\psi_{d}(1,2,3,4,5,6) \rangle ,
\label{ampl}
\ee
\be
{\bf r}  = \sum_{i=1}^{3}  {\bf r}_{i} -\sum_{i=4}^{6}  {\bf r}_{i}.
\label{r}
\ee
In \cite{Glozm} the expressions for the overlaps
$\psi_{Bp^{\prime}}^{d}({\bf k}_{p^{\prime}})$
were obtained for all $pB$ configurations produced by the antisymmetrization
of the $S$-wave component of the deuteron wave function. As the first step
to an appropriate choice of the RGM distribution function, $\chi({\bf r})$,
we use the conventional $NN$ deuteron wave function, $\chi_{NN}({\bf r})$,
consisting of the $S$- and $D$-components,
and modify it according to the standard RGM renormalization condition
\cite{Vildermuth,Oka}
\begin{equation}
\chi({\bf r})=\int\ \hat{N}^{-1/2}({\bf r,r}^{\prime})\chi_{NN}
({\bf r}^{\prime})d^{3}r^{\prime},
\label{renorm}
\end{equation}
where $\hat{N}({\bf r,r}^{\prime})$ -- is a norm operator
\begin{eqnarray}
\hat{N}({\bf r,r^{\prime}}) &\!\!\!=\!\!\!& \delta ({\bf r-r^{\prime}})
-9\langle\{\varphi_{N}(1,2,3)\varphi_{N}(4,5,6)\}_{ST=10}
\delta ({\bf r-r^{\prime\prime}})|
\nonumber \\
& & \times
\hat{P}_{36}|\{\varphi_{N}(1,2,3)
\varphi_{N}(4,5,6)\}_{ST=10}\delta ({\bf r^{\prime}-r^{\prime\prime}})\rangle .
\end{eqnarray}
Fig.2 displays the ratios
$\chi^{L=0}(r)/ \chi^{L=0}_{NN}(r)$ and $\chi^{L=2}(r)/ \chi^{L=2}_{NN}(r)$.
In the further calculations we approximate the radial parts of the
$S$- and $D$-components of the RGM distribution function by gaussians:
\begin{equation}
\frac{\chi^{L=0}(r)}{r}=\sum_{k}A_{k}e^{-\alpha_{k}r^{2}}, \ \ \
\frac{\chi^{L=2}(r)}{r}= r^2 \sum_{k}B_{k}e^{-\beta_{k}r^{2}}.
\label{approximation}
\end{equation}
As the nucleon and excited states wave functions
 we have used the wave functions
 of the Translationally Invariant Shell Model
\cite{Smirnov} for three particles,
i.e. the harmonic oscillator wave functions
 with oscillator parameter $b$, with the center-of-mass oscillations being
removed.
All detailes can be found in \cite{Glozm}.
The parameters of the approximation (\ref{approximation}) for the RGM
distribution function based on the Paris potential are given
in Table 1.
Making use the gaussian approximation all necessary calculations could be
performed analytically \cite{Glozm} and the Fourier transformation
(\ref{Fourier}) is given by
\begin{eqnarray}
\psi_{Bp^{\prime}}^{d}({\bf k}_{p^{\prime}}) &\!\!\!=\!\!\!& \chi
({\bf k}_{p^{\prime}}) \delta_{B,N}+3\sum_{m_{B},\mu_{B}}\left\langle
L_{B}S_{B}m_{B} \mu_{B}\left|\right. J_{B}M_{B}\right\rangle
\frac{1}{\sqrt{dim[f_{B}]}}\nonumber\\
&\!\!\!\times\!\!\! & \gamma^{X}_{B}\sum_{k}\ A_{k}\
I^{L_{B}}_{N_{B}L_{B},00}(k_{p^{\prime}};\alpha_{k})\
Y_{L_{B}m_{B}}^{\star}(\hat{\bf k}_{p^{\prime}})\
(-1)^{\frac{1}{2}+S_{B}+2T_{B}} \nonumber \\
& &  \times\sqrt{(2T_{B}+1)(2S_{B}+1)}
\langle S_{B}\frac{1}{2}\mu_{B}\mu_{p^{\prime}}\left|\right. 1M_{d}\rangle
\nonumber \\
& &  \times
\sum_{\stackrel{\scriptstyle S_{12}=T_{12}=0,1}{S_{45}=T_{45}=0,1}}
\langle [f_{B}]S_{B}T_{B}\left|\right. [2]S_{12}T_{12};\frac{1}{2}
\frac{1}{2}\rangle\nonumber \\
& &  \times (-1)^{S_{12}+S_{45}}\left\{
\begin{array}{ccc}
\frac{1}{2} & T_{12} & \frac{1}{2}\\
\frac{1}{2} & T_{45} & \frac{1}{2}\\
\end{array}\right\}\left\{
\begin{array}{ccc}
\frac{1}{2} & S_{12} & S_{B}\\
S_{45} & \frac{1}{2} & \frac{1}{2}\\
\frac{1}{2} & \frac{1}{2} & 1\\
\end{array}\right\}.
\label{general}
\end{eqnarray}
For the explanation of notations in (\ref{general}), see \cite{Glozm}.
The functions $I^{L_{B}}_{N_{B}L_{B},00}$
are summarized in Appendix A. The fractional parentage coefficients
(CFP), $\gamma_{B}^{X}$, for $N_{B}=3$ could be found in Appendix B
(the CFP for $N_{B}=0,\ 1,\ 2$ are given in \cite{Glozm}).The
physical meaning of the parameter $b$ appearing in these functions is
the m.s. radius of a nucleon quark core,
{\it i.e.} a characteristic distance
where quark exchanges becomes noticeable. In Table 2 we summarize the
baryons $B$ which are taken into account in our calculations and give the
effective numbers, $N_{Bp}^{d}$, for three values of the parameter $b$;
for the definition of the effective number $N_{Bp}^{d}$ see, {\it e.g.}
\cite{Glozm}. The baryons incorporated in the components
with the effective numbers
$N_{Bp}^{d}<10^{-5}$ were ignored.

It should be mentioned here that the baryons from lines
2-5 and 7-10 of Table 2, have the negative parity and thus produce
effective $P$-waves of the relative motion in the deuteron.
It was already mentioned
(\cite{Kuehn93} and \cite{Kob94}) that the $P$-waves should strongly affect
polarization phenomena of the reaction at the region of internal
momentum in the deuteron of few hundreds MeV/c.

There are at least two problems connected to the relativistic description
which we would like to discuss here shortly. First,
RGM is based on nonrelativistic dynamical equations. To establish a
correspondence between the observed momentum of the spectator and
the wave function argument we use the ``minimal relativization prescription''.
This means that we identify the light-cone variable $k$ with the relative
momentum of the $3q$ clusters in the deuteron, considering their dynamics
nonrelativistically, {\sl but} in terms of {\sl new variable} $k$
(see, {\it e.g.}, \cite{KobViz}).
Second, the different components of the deuteron wave function
($NN$, $NN^{\star}$ {\it etc.}) consists of ``particles'' ($3q$ clusters)
with different masses. In this case every component, involved in our
calculations, in the limit of infinite distances should
depend on its own ``internal momentum''.
However such a limit is allowed only for the usual n-p component
and the effective $NN^*$ components appear only in the nucleon overlap region
and they are direct consequence of the underlying quark structure of a baryon.
Thus this question cannot be solved at the baryon level. In the present work
we assume for simplicity that the "internal momenta" in the n-p
and $N-N^*$ chanels are the same.

\section{\sloppy Momentum distribution of the pro\-tons in the
deu\-te\-ron and po\-la\-ri\-za\-tion ob\-ser\-va\-bles
in the spec\-ta\-tor ap\-proxi\-ma\-tion}

According to the spectator approximation (\ref{me}) the reaction
matrix element squared is
\be
\sum_X|M_{dp\rightarrow p^{\prime}X} |^2=2\sigma_0\sum_B
\left|\psi^{d}_{Bp^{\prime}}(k_{p^{\prime}})\right|^{2}.
\label{sqared}
\ee
In (\ref{sqared}) it is assumed that
\be
\sum_{X}
\Gamma^{\dag}_{pB^{\prime}\rightarrow X}\
\Gamma_{pB\rightarrow X}=\sigma_0\delta_{BB^{\prime}},
\label{dig}
\ee
where $\sigma_0$ is of order of the total
$NN$ cross section. This means that the size of $B$ is supposed to
be the same as the proton's one and inelasticities in baryon-baryon
collisions are small at the energy scale of few GeV. Now the proton momentum
distribution in the deuteron is given by
\be
n(k_{p^{\prime}}) = \tilde{u}^{2}+\tilde{w}^{2}+u_{1}^{2}+v_{1}^{2}+
\frac{1}{2}v_{2}^{2}+v_{3}^{2}+\frac{1}{2}v_{4}^{2}+2v_{5}^{2}+
v_{6}^{2}+2v_{7}^{2}+v_{8}^{2},
\label{momdistrib}
\ee
where
\begin{eqnarray}
\tilde{u}  &\!\!\!=\!\!\!&  \chi^{L=0}(k_{p^{\prime}})-\frac{1}{9}\frac{1}
{\sqrt{2}}\sum_k\ A_{k}\ I_{00,00}^{0}(k_{p^{\prime}};\alpha_{k}),\\
\label{utilde}
\tilde{w} &\!\!\!=\!\!\!& \chi^{L=2}(k_{p^{\prime}}), \\
\label{wtilde}
u_1  &\!\!\!=\!\!\!& \frac{4}{9\sqrt{2}}\sum_{k}\ A_{k}\
I_{20,00}^{0}(k_{p^{\prime}};\alpha_{k}), \\
\label{u1}
v_1 &\!\!\!=\!\!\!&  v_2=v_5=v_6=\frac{4}{9}\sum_{k}\ A_{k}\
I_{11,00}^{1}(k_{p^{\prime}};\alpha_{k}), \\
\label{v1256}
v_3 &\!\!\!=\!\!\!& v_4=v_7=v_8=\frac{2}{9}\sqrt{3}
\sum_{k}\ A_{k}\ I_{31,00}^{1}(k_{p^{\prime}};\alpha_{k}).
\label{uwv}
\end{eqnarray}
The differential cross section of the reaction is proportional to
the proton momentum distribution, $n(k_{p^{\prime}})$, (see, {\em e.g.},
\cite{KobViz}) and the tensor analyzing power, $T_{20}$, and
polarization transfer coefficient, $\kappa_{0}$, are given by
\begin{eqnarray}
T_{20}  &\!\!\!=\!\!\!&  \frac{1}{\sqrt{2}}
\frac{2\sqrt{2}\tilde{u}\tilde{w}-\tilde{w}^{2}-\frac{1}{2}v_{2}^{2}
-\frac{1}{2}v_{4}^{2}+\frac{4}{5}v_{6}^{2}+\frac{4}{5}v_{8}^{2}}
{n(k_{p^{\prime}})},\\
\kappa_0  &\!\!\!=\!\!\!&
(\tilde{u}^{2}-\tilde{w}^{2}-\sqrt{\frac{1}{2}}\tilde{u}
\tilde{w}+u_{1}^{2}+v_{1}^{2}-\frac{1}{2}v_{2}^{2}+v_{3}^{2}-
\frac{1}{2}v_{4}^{2}+2v_{5}^{2}\nonumber\\
&\!\!\!-\!\!\!&\frac{1}{10}v_{6}^{2}+2v_{7}^{2}-
\frac{1}{10}v_{8}^{2})/n(k_{p^{\prime}}).
\label{t20kap}
\end{eqnarray}

\section{Numerical calculations and conclusions}

The results of our calculations and comparison with the experimental data
are shown in Figs.~3-9. Figs.~5 and 7 explore the dependence of the
polarization observables $T_{20}$ and $\kappa_{0}$ on the parameter $b$ of
the oscillator quark potential. In Figs.~3,~4,~6 and 8 we demonstrate
insensitivity of the current approach on the choice of input
$NN$ potential.

In  Fig.~9 we compare the results with the experimental data
on the $T_{20}-\kappa_{0}$ plane. It was recently shown
\cite{Kuehn93}, that a correlation between
$T_{20}$ and $\kappa_0$ should give important information about the non-nucleon
degrees of freedom in the deuteron. Particularly, in the case, when the
deuteron wave function consists of the two standard ($S-$ and $D-$wave)
components only, the parametric curve describing the correlation between
$T_{20}$ and $\kappa_{0}$ must lie on the circle. Our calculations give
an example of a deformation of this relation produced by additional
components of the deuteron wave function.

The points for the proton momentum distribution in the deuteron
$n(k_{p^{\prime}})$ (Figs.~3 and 4) were extracted from data for
the $p(d,p)$ reaction cross section \cite{Ableev92} (see also this
reference, as well as \cite{Ableev}, \cite{Punjabi} for $A(d,p)$ data).
$T_{20}$ was measured in \cite{Punjabi}-\cite{Aono} and
$\kappa_{0}$ in \cite{Nom,Cheng,Alpha,Anomalon}.

The conclusions of the present work are as follows:\\
(i) When the internal momentum in the deuteron is of a few hundreds MeV/c, the
effects of the quark exchange between three-quark clusters in the
deuteron are as important as the ones originating from the relative
motion of the nucleons (estimated by the standard two-nucleon potentials).\\
(ii) The $pN^{\star}$ components produce effective $P$-waves in the deuteron
which correct the momentum distribution of protons in the deuteron and
polarization observables of the deuteron breakup at high/ intermediate
energy in the ``right'' direction.\\
(iii) The results of the current approach depend weekly on the choice of
an input two-nucleon potential, but, at the same time, very sensitive to
the quark core radius of the nucleon, $b$. However, it should be noted the
value $b=0.8$~fm which provide the best fit to data is somewhat larger
than  the m.s.r. of the quark core of a free nucleon,
$b=0.5\div 0.6$~fm, commonly used in theoretical calculations.

It should be also stressed that we consider here the intermediate region
between the pure $NN$ and the perturbative QCD regimes
and our model cannot be used for a region of very high $k$.
\newpage

\noindent
{\bf Acknowledgement.} The authors express their thank to
E.~A.~Stro\-kov\-sky for
useful discussions. Two of authors (A.P.K. and A.I.S.) are grateful to
C.~F.~Perdrisat for sending the data for $T_{20}-\kappa_0$ correlation.
For two of us (A.P.K. and A.I.S.) this work was supported in part by grant
${\cal N}^o$ RFU000 of the International Science Foundation.
\vspace{1cm}

\section*{Appendix A}
\begin{eqnarray}
I^{0}_{00,00} &\!\!\! = \!\!\!& \left(\frac{27}{18+30\alpha_{k}b^{2}}
\right)^{\frac{3}{2}}b^{3}
\mbox{exp}\left\{-\frac{b^2k^{2}}{12}\frac{15+16\alpha_{k}b^{2}}
{3+5\alpha_{k}b^{2}}\right\}\nonumber \\
I^{1}_{11,00} &\!\!\! = \!\!\!& I^{0}_{00,00}\ g_{1}\ b\ k \nonumber \\
I^{0}_{20,00} &\!\!\! = \!\!\!& I^{0}_{00,00}\  \sqrt{6}
\left(g-\frac{3}{2}g_{1}g+
\frac{g_{1}^{2}}{2}b^{2}k^{2}\right)\nonumber \\
I^{2}_{22,00} &\!\!\! = \!\!\!& I^{0}_{00,00}\  \sqrt{\frac{3}{5}}g_{1}^2
b^2k^2\nonumber \\
I^{1}_{31,00} &\!\!\! = \!\!\!& I^{0}_{00,00}\  \sqrt{10}g_{1}\ b\ k
\left(g-\frac{3}{2}g_{1}g+\frac{3g_{1}^{2}}{10}b^{2}k^{2}\right),\nonumber
\end{eqnarray}
where
\begin{eqnarray}
g=\frac{4\alpha_{k}b^{2}-3}{15+16\alpha_{k}b^{2}},\ \ \
g_{1}=\frac{4\alpha_{k}b^{2}-3}{18+30\alpha_{k}b^{2}}.
\nonumber
\end{eqnarray}
\vskip 1cm

\noindent
{\large\bf Appendix B.}\\
\noindent
{\large \bf Baryon orbital wave functions
$|3(\lambda\mu)[f]L(r)\rangle$}
\vskip 0.5cm
\noindent
The choice of Jacobi coordinates is
\begin{eqnarray}
{\bf r}_{12}=\frac{{\bf r}_{1}-{\bf r}_{2}}{\sqrt{2}},\ \ \
\mbox{\boldmath $\rho$}=\frac{{\bf r}_{1}+{\bf r}_{2}-2{\bf r}_{3}}
{\sqrt{6}}.
\nonumber\end{eqnarray}
\vskip 0.3cm\noindent
\begin{eqnarray}
\hskip -0.2in
& & \hskip -2.0in |3(30)[3]1(111)\rangle = \frac{1}{2}\sqrt{\frac{5}{3}}
\varphi_{200}
({\bf r}_{12})\varphi_{11M}(\mbox{\boldmath $\rho$})+\frac{1}{\sqrt{3}}
\sum_{m_{1},m_{2}}\langle21\ m_{1}m_{2}\left|\right. 1M\rangle\nonumber\\
 & & \times\varphi_{22m_{1}}({\bf r}_{12})
\varphi_{11m_{2}}(\mbox{\boldmath $\rho$})-\frac{1}{2}\varphi_{000}
({\bf r}_{12})\varphi_{31M}(\mbox{\boldmath $\rho$})\nonumber\\
& & \hskip -2.0in|3(30)[21]1(112)\rangle = -\frac{\sqrt{5}}{6}\varphi_{200}
({\bf r}_{12})\varphi_{11M}(\mbox{\boldmath $\rho$})-\frac{1}{3}
\sum_{m_{1},m_{2}}\langle 21\ m_{1}m_{2}\left|\right. 1M\rangle\nonumber\\
 & & \times\varphi_{22m_{1}}({\bf r}_{12})
\varphi_{11m_{2}}(\mbox{\boldmath $\rho$})-\frac{\sqrt{3}}{2}\varphi_{000}
({\bf r}_{12})\varphi_{31M}(\mbox{\boldmath $\rho$})\nonumber\\
|3(30)[21]1(121)\rangle &\!\!\! = \!\!\!&
\frac{\sqrt{3}}{2}\varphi_{31M}({\bf r}_{12})
\varphi_{000}(\mbox{\boldmath $\rho$}) -\frac{\sqrt{5}}{6}\varphi_{11M}
({\bf r}_{12})\varphi_{200}(\mbox{\boldmath $\rho$})\nonumber\\
&\!\!\!\!-\!\!\!\!&\frac{1}{3}\sum_{m_{1},m_{2}}\langle 12\
m_{1}m_{2}\left|\right. 1M
\rangle\varphi_{11m_{1}}({\bf r}_{12})
\varphi_{22m_{2}}(\mbox{\boldmath $\rho$})\nonumber\\
|3(11)[21]1(112)\rangle &\!\!\! = \!\!\!& \frac{2}{3}\varphi_{200}
({\bf r}_{12})\varphi_{11M}(\mbox{\boldmath $\rho$})-\frac{\sqrt{5}}{3}
\sum_{m_{1},m_{2}}\langle 21\ m_{1}m_{2}\left|\right. 1M\rangle\nonumber\\
 & & \times\varphi_{22m_{1}}({\bf r}_{12})\varphi_{11m_{2}}
 (\mbox{\boldmath $\rho$}) \nonumber\\
|3(11)[21]1(121)\rangle &\!\!\! = \!\!\!& \frac{2}{3}\varphi_{11M}
({\bf r}_{12})\varphi_{200} (\mbox{\boldmath $\rho$})-\frac{\sqrt{5}}{3}
\sum_{m_{1},m_{2}}\langle 12\ m_{1}m_{2}\left|\right. 1M\rangle\nonumber\\
 & &\times\varphi_{11m_{1}}({\bf r}_{12})\varphi_{22m_{2}}
 (\mbox{\boldmath $\rho$}) \nonumber\\
|3(30)[1^3]1(123)\rangle &\!\!\! = \!\!\!& \frac{1}{2}\varphi_{31M}
({\bf r}_{12})\varphi_{000}(\mbox{\boldmath $\rho$})-\frac{1}{2}
\sqrt{\frac{5}{3}}
\varphi_{11M}({\bf r}_{12})\varphi_{200}(\mbox{\boldmath $\rho$})\nonumber\\
 & & -\frac{1}{\sqrt{3}}
\sum_{m_{1},m_{2}}\langle 12\ m_{1}m_{2}\left|\right. 1M\rangle
\varphi_{11m_{1}}({\bf r}_{12})\varphi_{22m_{2}}(\mbox{\boldmath $\rho$})
 \nonumber\\
|3(11)[21]2(112)\rangle  &\!\!\!=\!\!\!&  -\sum_{m_{1},m_{2}}
\langle 21\ m_{1}m_{2}\left|\right. 2M\rangle\varphi_{22m_{1}}({\bf r}_{12})
\varphi_{11m_{2}} (\mbox{\boldmath $\rho$}) \nonumber\\
|3(11)[21]2(121)\rangle  &\!\!\!=\!\!\!&  \sum_{m_{1},m_{2}}
\langle 12\ m_{1}m_{2}\left|\right. 2M\rangle\varphi_{11m_{1}}({\bf r}_{12})
\varphi_{22m_{2}} (\mbox{\boldmath $\rho$}) \nonumber\\
|3(30)[3]3(111)\rangle &\!\!\! = \!\!\!& \frac{\sqrt{3}}{2}
\sum_{m_{1},m_{2}}\langle 21\ m_{1}m_{2}\left|\right. 3M\rangle
\varphi_{22m_{1}}({\bf r}_{12})\varphi_{11m_{2}}
(\mbox{\boldmath $\rho$})\nonumber \\
& & \hspace{0.5in} -\frac{1}{2}\varphi_{000}({\bf r}_{12})\varphi_{33M}
(\mbox{\boldmath $\rho$})
\nonumber\\
|3(30)[21]3(112)\rangle  &\!\!\!=\!\!\!&  -\frac{1}{2}\sum_{m_{1}m_{2}}
\langle 21\ m_{1}m_{2}\left|\right. 3M\rangle \varphi_{22m_{1}}({\bf r}_{12})
\varphi_{11m_{2}}(\mbox{\boldmath $\rho$})\nonumber \\
& & \hspace{0.5in} -\sqrt{\frac{3}{4}}
\varphi_{000}({\bf r}_{12})\varphi_{33M}(\mbox{\boldmath $\rho$}) \nonumber\\
|3(30)[21]3(121)\rangle  &\!\!\!=\!\!\!&  -\sqrt{\frac{3}{4}}
\varphi_{33M}({\bf r}_{12})\varphi_{000}(\mbox{\boldmath $\rho$})\nonumber \\
& &  -\frac{1}{2}\sum_{m_{1}m_{2}}
\langle 12\ m_{1}m_{2}\left|\right. 3M\rangle \varphi_{11m_{1}}({\bf r}_{12})
\varphi_{22m_{2}}(\mbox{\boldmath $\rho$})\nonumber\\
|3(30)[1^3]3(123)\rangle &\!\!\! = \!\!\!& \frac{1}{2}\varphi_{33M}
({\bf r}_{12})\varphi_{000}(\mbox{\boldmath $\rho$})\nonumber \\
& & -\frac{\sqrt{3}}{2}
\sum_{m_{1},m_{2}}\langle 12\ m_{1}m_{2}\left|\right. 3M\rangle
\varphi_{11m_{1}}({\bf r}_{12})\varphi_{22m_{2}}(\mbox{\boldmath $\rho$})
 \nonumber
\end{eqnarray}
\newpage

\hfill {\it Table 1.}

\begin{center}
{\bf The parameters of the approximation (\ref{approximation}) for the RGM
distribution function based on the Paris potential. The oscillator parameter
$b=0.8\ fm$.}
\end{center}
\begin{center}
\begin{tabular}{||c|c|c|c|c||}  \hline
$n$ & $A_{n}$ ($fm^{-3/2}$) & $\alpha_{n}$ ($fm^{-2}$) &
 $B_{n}$ ($fm^{-7/2}$) & $\beta_{n}$ ($fm^{-2}$) \\ \hline
 1 & $1.3863\cdot 10^{-2}$ & $8.845\cdot 10^{-3} $& $3.1946\cdot 10^{-4}$&
 $4.1231\cdot 10^{-2}$
  \\ \hline
 2 & $7.1097\cdot 10^{-2} $& $3.1668\cdot 10^{-2}$& $ 3.6241\cdot 10^{-3}$&
 $1.1980\cdot 10^{-1}
$ \\ \hline
 3 & $1.7493\cdot 10^{-1} $& $1.0619\cdot 10^{-1}$& $  1.7823\cdot 10^{-2}$&
 $2.6855\cdot 10^{-1}$ \\ \hline
 4 & $2.9944\cdot 10^{-1} $& $3.8459\cdot 10^{-1}$& $ 7.5863\cdot 10^{-2} $&
 $5.5486\cdot 10^{-1}$
 \\ \hline
 5 & $-6.8075\cdot 10^{-1}$& $2.4318             $& $  3.3025\cdot 10^{-1}$&
 $1.3350 $ \\ \hline
 6 & $1.8345\cdot 10^{-1} $& $3.5360             $& $-2.6708\cdot 10^{-1}$&
 $5.4722             $ \\ \hline
\end{tabular}
\end{center}

\hfill {\it Table 2.}

\begin{center}
{\bf The effective numbers $N_{Bp}^{d}$ of different baryon-proton
configurations in the deuteron. The calculations were done for the
Paris potential.}
\end{center}
\begin{center}
\begin{tabular}{||c|c|c|c|c|c||}  \hline
 &  $_{_{\ds J^{P}T}}$    &$_{_{\mbox{State}}}$  &
\multicolumn{3}{c||}{$N_{Bp}^{d}$}\\ \cline{4-6}
 &                    &    & $b=0.5\ fm$ & $b=0.7\ fm$ & $b=0.8\ fm$
 \\ \hline\hline
 $^{^{\ds 1}}$  &  $^{\ds \frac{1}{2}^+\ \frac{1}{2}}$ &
 $^{\ds |0(00)[3]0\frac{1}{2}\frac{1}{2}\rangle}$  & $^{^{\ds 0.982}}$
 &$^{^{\ds 0.996}}$&$^{^{\ds 1.005}}$
 \\ \hline
 $^{^{\ds 2}}$  &  $^{\ds \frac{3}{2}^-\ \frac{1}{2}}$ &
 $^{\ds |1(10)[21]1\frac{1}{2}\frac{1}{2}\rangle}$ &
 $^{^{\ds 1.52\times 10^{-3}}}$ &$^{^{\ds 3.60\times 10^{-3}}}$&
 $^{^{\ds 4.46\times 10^{-3}}}$ \\ \hline
 $^{^{\ds 3}}$  &  $^{\ds \frac{1}{2}^-\ \frac{1}{2}}$ &
 $^{\ds |1(10)[21]1\frac{1}{2}\frac{1}{2}\rangle}$ & $^{^{\ds 7.59\times
 10^{-4}}}$
 &$^{^{\ds 1.80\times 10^{-3}}}$&$^{^{\ds 2.23\times 10^{-3}}}$\\ \hline
 $^{^{\ds 4}}$  &  $^{\ds \frac{1}{2}^-\ \frac{1}{2}}$ &
 $^{\ds |1(10)[21]1\frac{3}{2}\frac{1}{2}\rangle}$ &
 $^{^{\ds 3.79\times 10^{-4}}}$&$^{^{\ds 9.01\times 10^{-4}}}$&
 $^{^{\ds 1.16\times 10^{-3}}}$ \\ \hline
 $^{^{\ds 5}}$  &  $^{\ds \frac{3}{2}^-\ \frac{1}{2}}$ &
 $^{\ds |1(10)[21]1\frac{3}{2}\frac{1}{2}\rangle}$ &
 $^{^{\ds 7.59\times 10^{-4}}}$&$^{^{\ds 1.80\times 10^{-3}}}$&
 $^{^{\ds 2.23\times 10^{-3}}}$\\ \hline
 $^{^{\ds 6}}$  &  $^{\ds \frac{1}{2}^+\ \frac{1}{2}}$ &
 $^{\ds |2(20)[21]0\frac{1}{2}\frac{1}{2}\rangle}$ &
 $^{^{\ds 2.32\times 10^{-3}}}$&$^{^{\ds 5.47\times 10^{-3}}}$&
 $^{^{\ds 6.75\times 10^{-3}}}$\\ \hline
 $^{^{\ds 7}}$  &  $^{\ds \frac{1}{2}^-\ \frac{1}{2}}$ &
 $^{\ds |3(30)[21]1\frac{1}{2}\frac{1}{2}\rangle}$ &
 $^{^{\ds 2.52\times 10^{-4}}}$&$^{^{\ds 3.63\times 10^{-4}}}$&
 $^{^{\ds 3.67\times 10^{-4}}}$\\ \hline
 $^{^{\ds 8}}$  &  $^{\ds \frac{3}{2}^-\ \frac{1}{2}}$ &
 $^{\ds |3(30)[21]1\frac{1}{2}\frac{1}{2}\rangle}$ &
 $^{^{\ds 5.04\times 10^{-4}}}$&$^{^{\ds 7.25\times 10^{-4}}}$&
 $^{^{\ds 7.33\times 10^{-4}}}$\\ \hline
 $^{^{\ds 9}}$  &  $^{\ds \frac{1}{2}^-\ \frac{1}{2}}$ &
 $^{\ds |3(30)[21]1\frac{3}{2}\frac{1}{2}\rangle}$ &
 $^{^{\ds 1.26\times 10^{-4}}}$&$^{^{\ds 1.81\times 10^{-4}}}$&
 $^{^{\ds 1.83\times 10^{-4}}}$\\ \hline
 $^{^{\ds 10}}$ &  $^{\ds \frac{3}{2}^-\ \frac{1}{2}}$ &
 $^{\ds |3(30)[21]1\frac{3}{2}\frac{1}{2}\rangle}$ &
 $^{^{\ds 2.52\times 10^{-4}}}$&$^{^{\ds 3.63\times 10^{-4}}}$&
 $^{^{\ds 3.67\times 10^{-4}}}$\\ \hline
\end{tabular}
\end{center}

\newpage

\noindent
Fig.~2. Ratios $\chi^{L=0}(r)/ \chi^{L=0}_{NN}(r)$ and
$\chi^{L=2}(r)/ \chi^{L=2}_{NN}(r)$. Calculations are performed
with the Paris potential and the parameter $b=0.8$ fm.
\vskip 0.5cm
\noindent
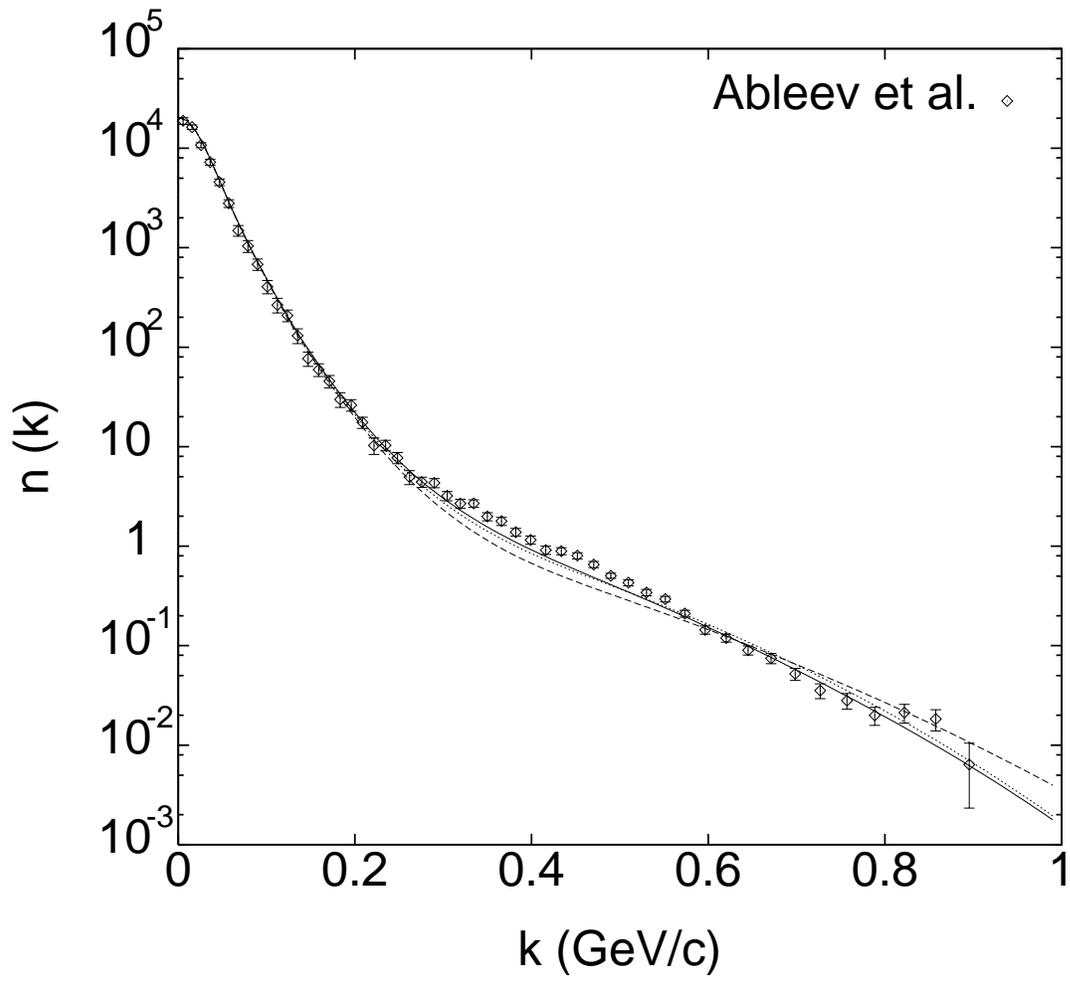
Fig.~3. Momentum distribution of the proton in the deuteron. Solid line
stands for calculations within the current approach with $b=0.8$ fm,
dashed and short-dashed lines -- to $b=0.5$ and $0.7$ fm,
respectively. Calculations are performed with the Paris potential.
The points are extracted from $^1H(d,p)$ cross section data \cite{Ableev92}.
\vskip 0.5cm
\noindent
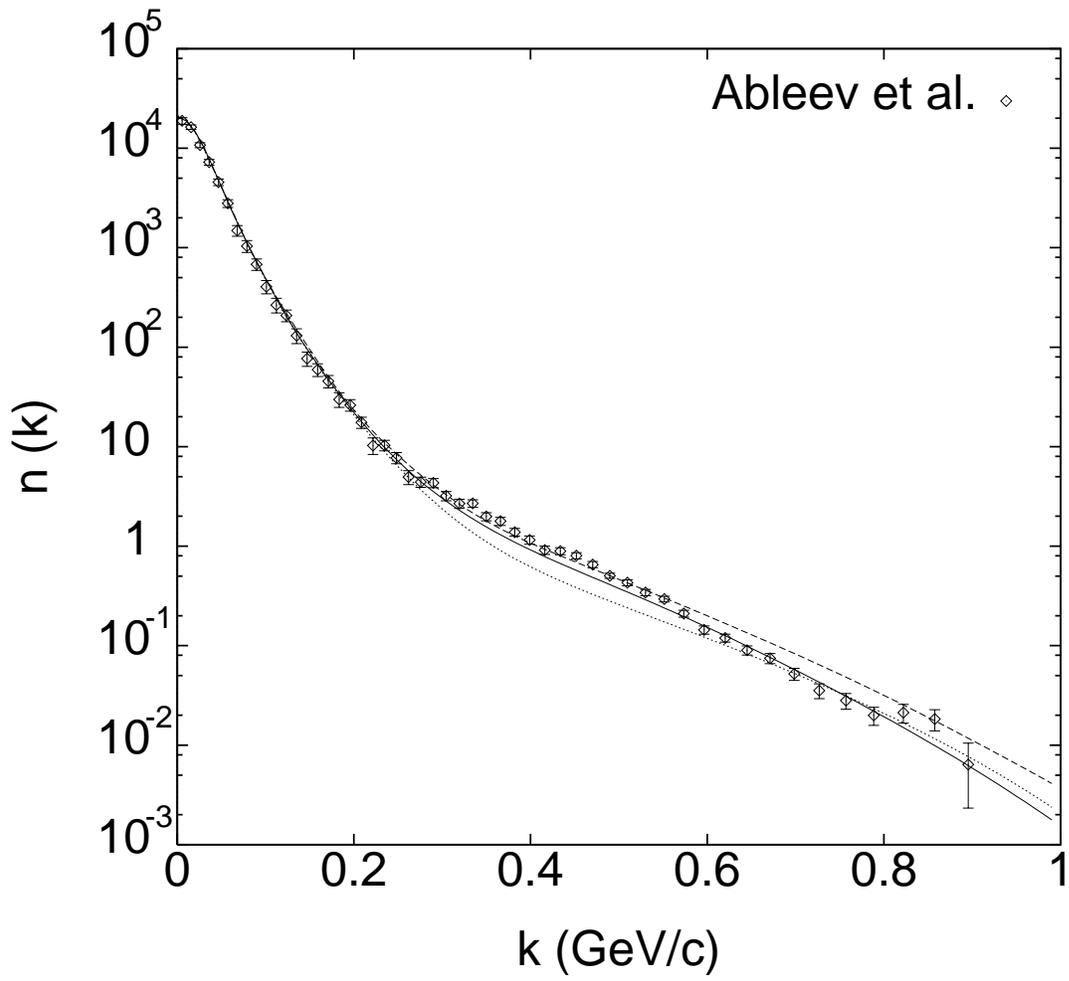
Fig.~4. Momentum distribution of the proton in the deuteron. Solid and
dashed lines stand for calculations within the current approach with
$b=0.8$ fm, with Paris and Reid-soft-core potentials, respectively.
By short-dashed line we denote the curve obtained in the IA with
Paris potential.
\vskip 0.5cm
\noindent
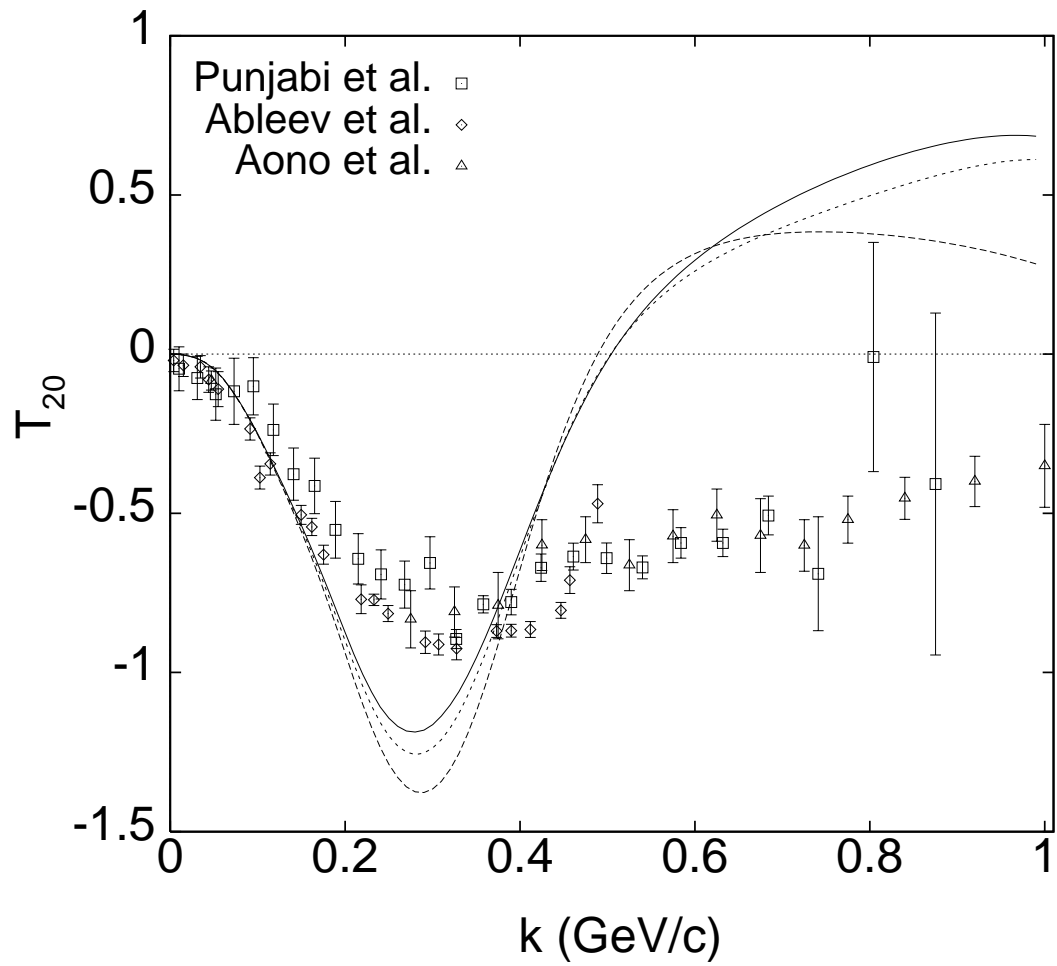
Fig.~5. $T_{20}$ in the current approach. Solid line corresponds to
$b=0.8$ fm, dashed and short-dashed lines -- to $b=0.7$ and $0.5$ fm,
respectively. Calculations are performed with the Paris potential.
The experimental data are from \cite{Punjabi}-\cite{T20}.
\vskip 0.5cm
\noindent
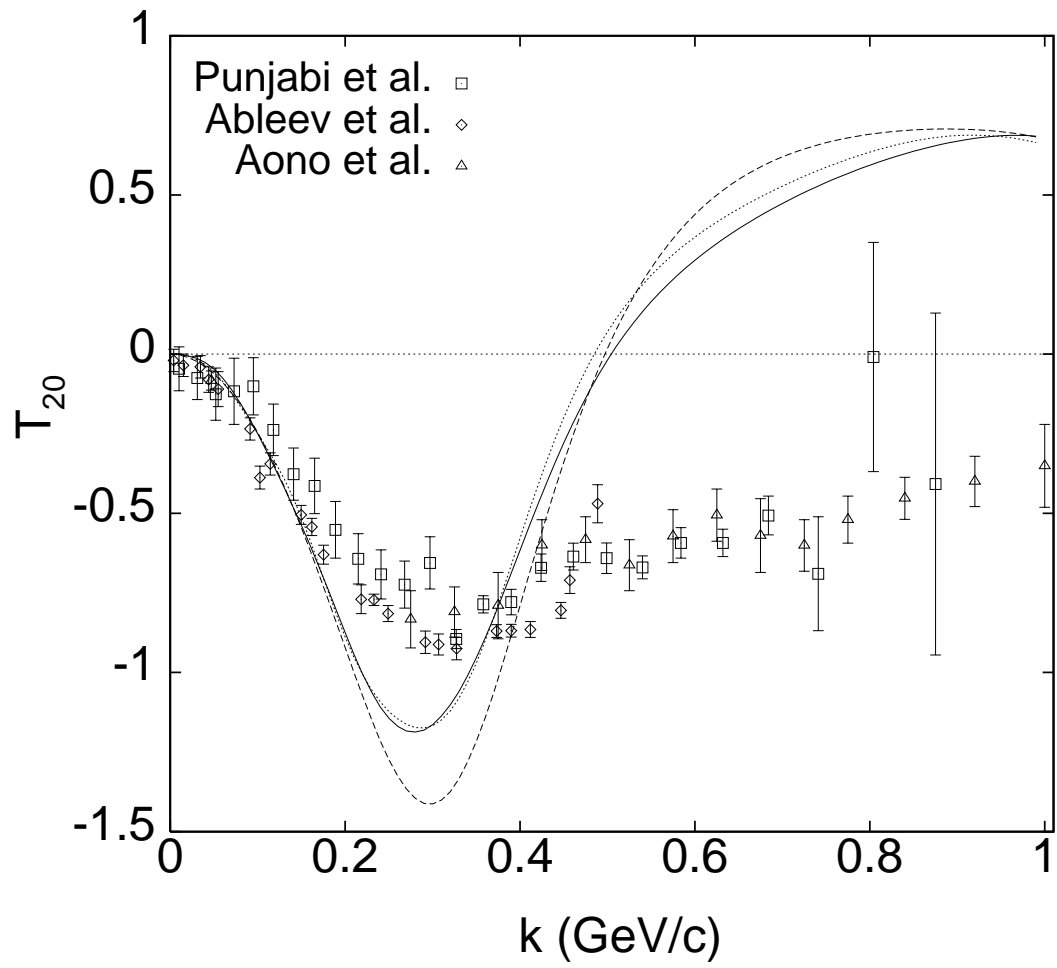
Fig.~6. $T_{20}$ in the current approach ($b=0.8$ fm), with the Paris
potential (solid line) and the Reid-soft-core potential (short-dashed line).
Dashed line stands for the IA calculations with the Paris potential.
The experimental data are from \cite{Punjabi}-\cite{T20}.
\vskip 0.5cm
\noindent
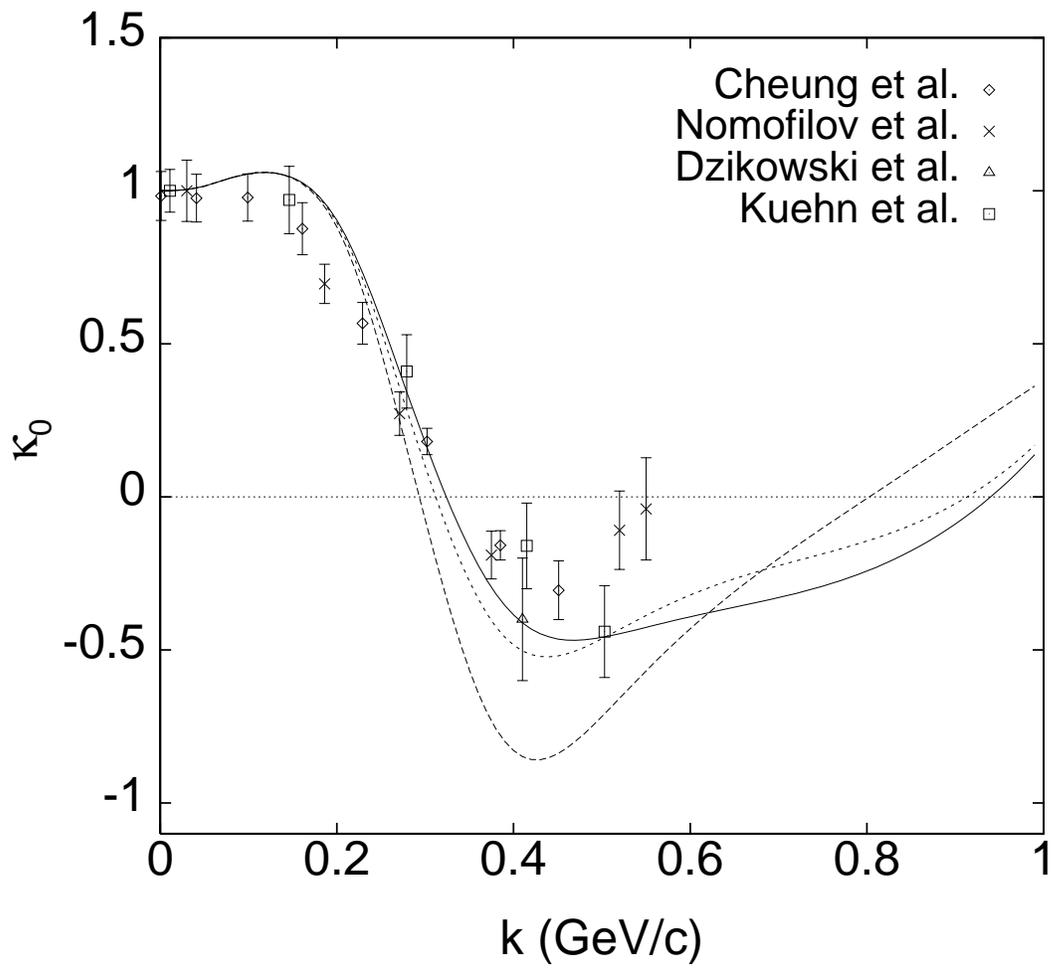
Fig.~7. The same as in Fig.~5, for $\kappa_{0}$. The experimental data
are from \cite{Nom,Cheng,Alpha}.
\vskip 0.5cm
\noindent
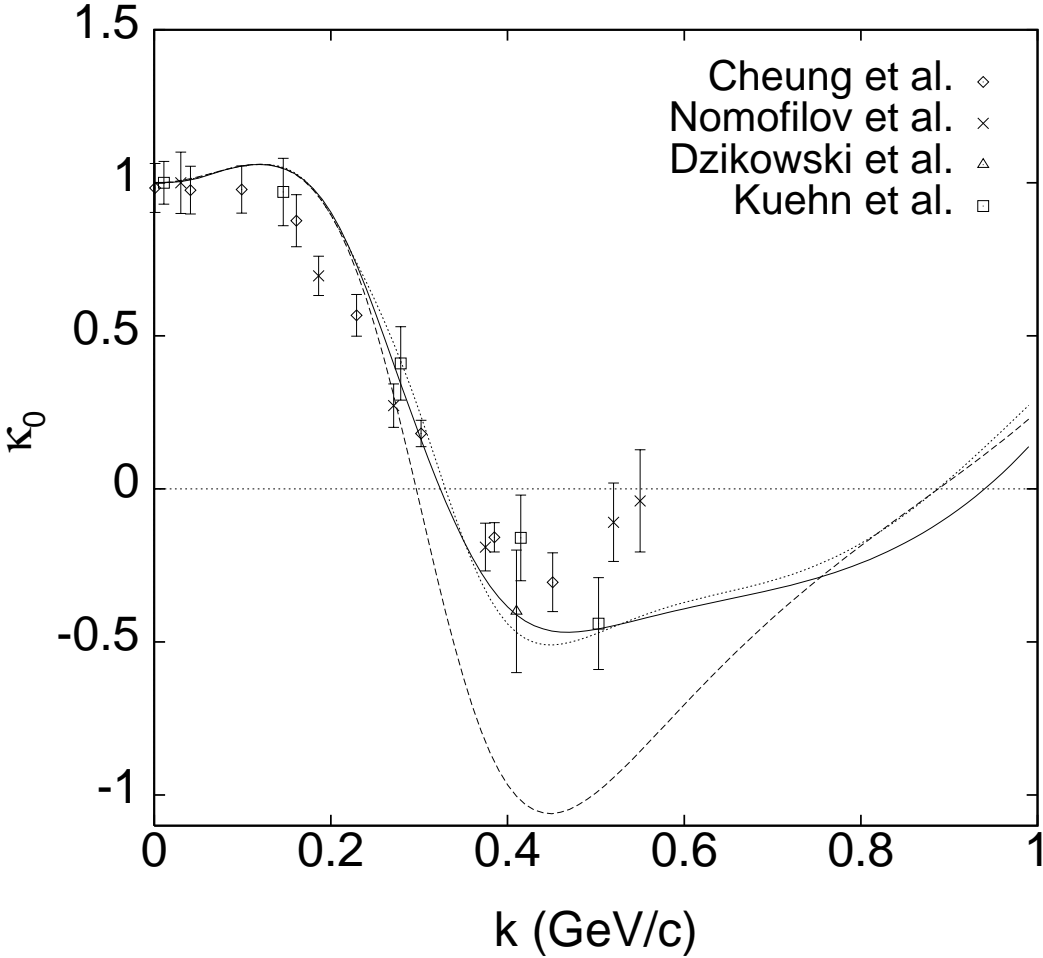
Fig.~8. The same as in Fig.~6, for $\kappa_{0}$.
\vskip 0.5cm
\noindent
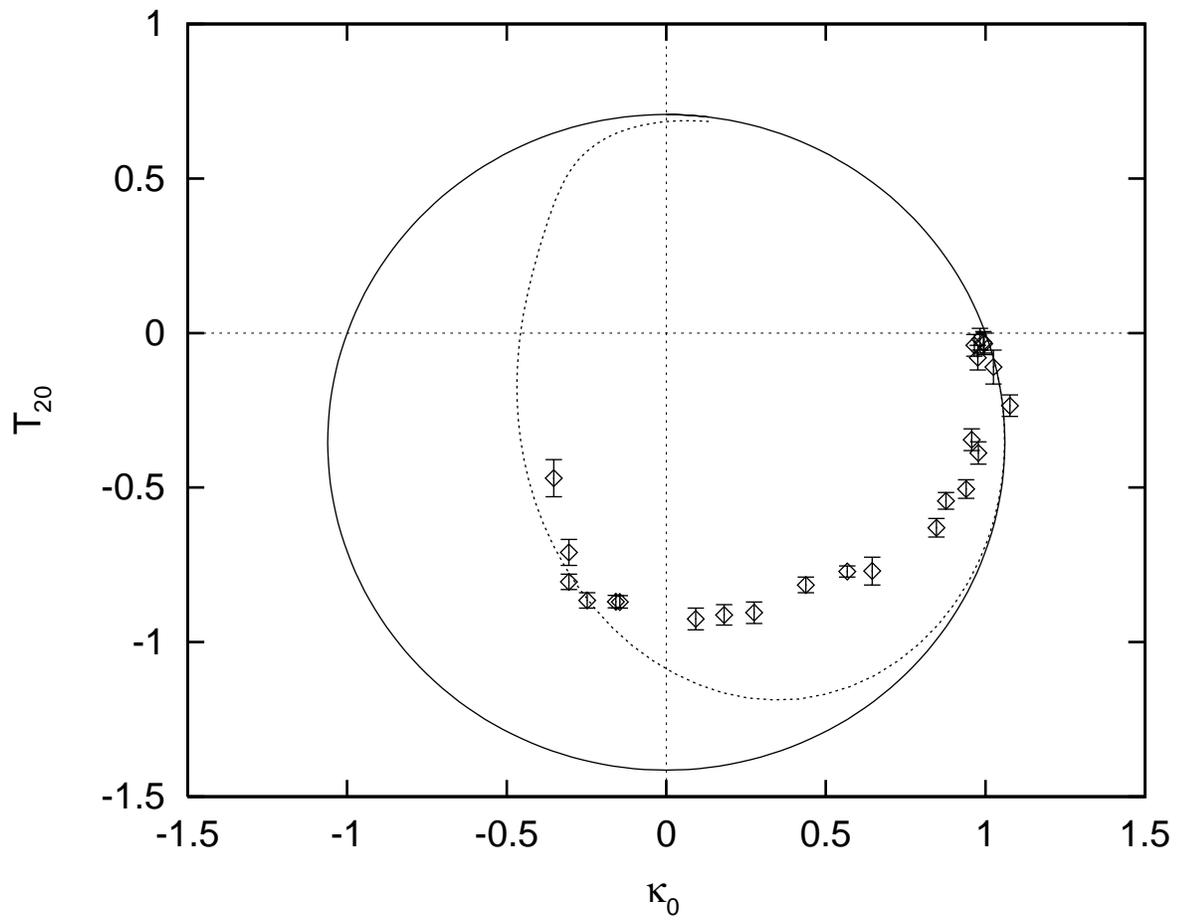
Fig.~9. The $\kappa_{0}\ \mbox{vs}\ T_{20}$ plot in the current approach
(dashed line). The solid line represents the results of IA \cite{Kuehn93}.
Calculations are performed with the Paris potential.
Data are from \cite{Cheng}.
\end{document}